# MDR-DeePC: Model-Inspired Distributionally Robust Data-Enabled Predictive Control

Shihao Li*, Jiachen Li*, Christopher Martin*, Soovadeep Bakshi*, Dongmei Chen*

* The University of Texas at Austin, Austin, TX 78721 USA (email: shihaoli01301@utexas.edu, jiachenli@utexas.edu, cbmartin129@utexas.edu, soovadeep.bakshi@utexas.edu, dmchen@me.utexas.edu)

**Abstract**: This paper presents a Model-Inspired Distributionally Robust Data-enabled Predictive Control (MDR-DeePC) framework for systems with partially known and uncertain dynamics. The proposed method integrates model-based equality constraints for known dynamics with a Hankel matrix-based representation of unknown dynamics. A distributionally robust optimization problem is formulated to account for parametric uncertainty and stochastic disturbances. Simulation results on a triple-mass-spring-damper system demonstrate improved disturbance rejection, reduced output oscillations, and lower control cost compared to standard DeePC. The results validate the robustness and effectiveness of MDR-DeePC, with potential for real-time implementation pending further benchmarking.

## 1. INTRODUCTION

Modern control systems in safety-critical applications, such as aerospace, medical devices, and energy systems, demand robust strategies to manage uncertainties. Traditional model-based methods like Model Predictive Control (MPC) depend on precise system models, which are challenging to obtain due to nonlinearities, time-varying parameters, and unmodeled dynamics, leading to potential performance issues or instability (Garcia, Prett, and Morari, 1989).

As an alternative, data-driven methods like Data-enabled Predictive Control (DeePC) use historical data to predict and control system behavior without explicit models, proving effective for complex systems (Hewing et al., 2020). However, DeePC often lacks robustness guarantees against uncertainties such as noise or parameter variations. In response, distributionally robust optimization (DRO) has been proposed as a means to design controllers that minimize worst-case costs across uncertainty distributions, balancing conservatism and performance (Delage and Ye, 2010). While DRO enhances robustness in data-driven settings, it does not inherently address the limitations arising from the absence of structural model information, motivating the development of hybrid control strategies.

Hybrid control approaches have emerged to combine the strengths of both model-based and data-driven strategies. Watson (2025) proposed Hybrid Data-enabled Predictive Control (HDeePC) to incorporate partial model knowledge into DeePC, thereby reducing data dependency and computational complexity. This hybrid strategy aligns with emerging methodologies utilizing Gaussian processes (Hewing et al., 2019) or neural networks (Maddalena et al., 2022), and finds additional theoretical support in Koopman-based hybrid control methods (Korda and Mezicé, 2018). However, the method developed by Watson does not consider robustness against noise and uncertainties.

Recent research has further advanced data-driven control by introducing regularization techniques to manage noise (Berberich et al., 2023) and extending DeePC to nonlinear systems using DRO for stochastic uncertainties (Coulson et al., 2019a, 2019b; Dörfler et al., 2022). In parallel to these recent developments in data-driven control, hybrid control methods have recently been applied in various applications, such as hybrid MPC for building energy systems (Stoffel et al., 2022) and mixed-integer hybrid MPC for behavioral interventions (Dong et al., 2022). However, many existing methods are tailored to specific use cases and do not provide a unified robust optimization framework. For instance, the approach developed by Coulson focuses solely on stochastic uncertainty (Coulson et al., 2019b). A more generalized consideration on robust control is desired.

This paper presents the Model-Inspired Distributionally Robust DeePC (MDR-DeePC) framework, which integrates robust optimization into the HDeePC architecture to address both parametric uncertainties in known dynamics and stochastic disturbances in unknown dynamics. By incorporating partial model knowledge within a data-driven predictive control framework, MDR-DeePC improves robustness, reduces dependency on large datasets, and enhances computational efficiency. The proposed method is validated through a case study on a triple-mass-spring-damper system, demonstrating improved disturbance rejection, reduced output oscillations, and lower control cost compared to standard DeePC. The paper is organized as follows: Section 2 presents the problem formulation and system model; Section 3 details the MDR-DeePC framework; Section 4 describes the case study and simulation setup, and results; and Section 5 concludes the paper and outlines future research directions.

## 2. Problem Formulation

In this section, the development of a Model-Inspired Distributionally Robust Data-enabled Predictive Control (MDR-DeePC) strategy is presented. The approach merges model-based and data-driven control methods to ensure reliable performance under uncertain conditions, with a focus on safety-critical systems. Additionally, the system is partitioned into known dynamics (based on models) and unknown dynamics (driven by data). By applying behavioral theory, data-driven representations are incorporated, and a hybrid distributionally robust optimization framework is established to address uncertainties effectively.

*2.1 System Model with Hybrid Decomposition*

Consider a discrete-time linear system characterized by a state vector $x_k \in \mathbb{R}^n$, control input $u_k \in \mathbb{R}^m$, and measured output $y_k \in \mathbb{R}^p$. The state vector is decomposed into two components: the known part $x_{k,k} \in \mathbb{R}^{n_k}$, which is governed by an explicitly modeled dynamic, and the unknown part $x_{u,k} \in \mathbb{R}^{n_u}$, whose behavior is captured from data. Thus, the overall state dimension satisfies $n = n_k + n_u$. The dynamics of the system are modeled as follows:

$$x_{k+1} = \begin{bmatrix} x_{k,k+1} \\ x_{u,k+1} \end{bmatrix} = \begin{bmatrix} A_k & A_{ku} \\ A_{uk} & A_u \end{bmatrix} \begin{bmatrix} x_{k,k} \\ x_{u,k} \end{bmatrix} + \begin{bmatrix} B_k \\ B_u \end{bmatrix} u_k + w_k \quad (1)$$

The block matrix $\begin{bmatrix} A_k & A_{ku} \\ A_{uk} & A_u \end{bmatrix}$ models how each subsystem evolves in time, with $A_k$ and $A_u$ capturing self-dynamics, and $A_{ku}$, $A_{uk}$ representing coupling between the known and unknown parts. The block vector $\begin{bmatrix} B_k \\ B_u \end{bmatrix}$ reflects how the input $u_k$ influences both subsystems. Finally, the disturbance term $w_k$ includes any additional model uncertainties or noise that affects the overall state.

The system output $y_k$ is decomposed to distinguish between modeled and data-driven dynamics. The modeled output $y_{k,k}$ is generated solely from the known state $x_{k,k}$ and input $u_k$, while the data-driven output $y_{u,k}$ may depend on both the known and unknown states. Under the assumption that the direct feedthrough from the unknown state to the modeled output is negligible, the output equation is expressed as:

$$y_k = \begin{bmatrix} y_{k,k} \\ y_{u,k} \end{bmatrix} = \begin{bmatrix} C_k(\theta) & 0 \\ C_{uk} & C_u \end{bmatrix} \begin{bmatrix} x_{k,k} \\ x_{u,k} \end{bmatrix} + \begin{bmatrix} D_k \\ D_u \end{bmatrix} u_k + v_k \quad (2)$$

where $y_{k,k} \in \mathbb{R}^{p_k}$ corresponds to the known dynamics and $y_{u,k} \in \mathbb{R}^{p_u}$ corresponds to the unknown dynamics, so that $p = p_k + p_u$.

In our work, this hybrid partitioning enables us to merge the advantages of a model-based representation for the known dynamics with data-driven characterizations for the unknown dynamics, thereby forming the basis for our robust extension.

*2.2 Uncertainties in Model and Data*

The system is influenced by two categories of uncertainties:

**(1) Parametric Uncertainties (Model-Based)**: The matrices $A_k(\theta)$, $A_{ku}(\theta)$, $A_{uk}(\theta)$, $B_k(\theta)$, and $C_k(\theta)$ are dependent on $\theta$, which resides within a predefined set $\Theta$ (e.g., polytopic or norm-bounded). The control strategy must remain robust to all potential variations of $\theta \in \Theta$, achieved through robust optimization techniques. In contrast to (Coulson et al. 2019b), who concentrate exclusively on stochastic uncertainties in a fully data-driven framework, our method explicitly incorporates parametric uncertainties in the known model, thereby improving its applicability to systems with partially known dynamics.

**(2) Stochastic Uncertainties (Disturbances and Data Noise)**: The disturbances $w_k$ and $v_k$ exhibit stochastic behavior with unknown distributions but are confined within ambiguity sets $\mathcal{P}_w$ and $\mathcal{P}_v$, respectively. These sets are derived from data and prior knowledge (e.g., using Wasserstein distance or moment constraints), akin to the methodology in Coulson et al. (2019b). The input-output data for the unknown subsystem is subject to noise, introducing additional uncertainty that is managed using a distributionally robust approach. By distinguishing between deterministic uncertainties ($\theta$) and stochastic uncertainties ($w_k, v_k$), separate control strategies can be developed, and together they can enhance the system robustness beyond those achieved by purely data-driven methods.

*2.3 Data-Driven Characterization of Unknown Dynamics*

Historical input-output data are utilized to represent the dynamics of the subsystem for which a model is not available. Provided that the experiment signal $u_d$ is sufficiently rich, the collected data can be arranged into a structured matrix known as a Hankel matrix, in which each ascending anti-diagonal (from left to right) contains identical elements. Specifically, given a sequence of vectors $s = [s_1, s_2, \ldots, s_T]$, where each $s_i \in \mathbb{R}^m$ represents a data point at time $i$, the Hankel matrix of depth $L$ is explicitly defined as:

$$H_L(s) = \begin{bmatrix} s_1 & s_2 & s_3 & \cdots & s_{T-L+1} \\ s_2 & s_3 & s_4 & \cdots & s_{T-L+2} \\ s_3 & s_4 & s_5 & \cdots & s_{T-L+3} \\ \vdots & \vdots & \vdots & \ddots & \vdots \\ s_L & s_{L+1} & s_{L+2} & \cdots & s_T \end{bmatrix} \quad (3)$$

This matrix has $L$ rows and $T - L + 1$ columns, with the property that each element satisfies $H_L(s)_{i,j} = s_{i+j-1}$; in other words, it captures a sliding window of $L$ consecutive elements from the sequence $s$, with each column representing a one-step shift.

In the proposed approach, Hankel matrices are constructed from the offline input and output data sequences. Let $u_d = [u_1, u_2, \ldots, u_T]$ be the sequence of input vectors, and $y_d^u = [y_1^u, y_2^u, \ldots, y_T^u]$ be the corresponding sequence of output vectors from the unknown subsystem. The Hankel matrix constructed from $u_d$ with depth $T_{\text{ini}} + N$ (with $T_{\text{ini}}$ as the initial horizon and $N$ as the prediction horizon) is denoted as $H_{T_{\text{ini}}+N}(u_d)$, and the corresponding Hankel matrix of the same depth constructed from $y_d^u$ is denoted as $H_{T_{\text{ini}}+N}(y_d^u)$.

Using these matrices, the future behavior of the unknown subsystem is parameterized by a coefficient vector $g$ according to:

$$\begin{bmatrix} H_{T_{\text{ini}}+N}(u_d) \\ H_{T_{\text{ini}}+N}(y_d^u) \end{bmatrix} g = \begin{bmatrix} u_{\text{ini}} \\ y_{\text{ini}} \\ u \\ y \end{bmatrix}. \quad (4)$$

In Eq. (4), $H_{T_{\text{ini}}+N}(u_d)$ is the Hankel matrix formed from the offline input sequence $u_d$, and $H_{T_{\text{ini}}+N}(y_d^u)$ is constructed from the offline output data $y_d^u$ that corresponds only to the unknown subsystem. The vector $u_{\text{ini}}$ consists of the most recent $T_{\text{ini}}$ input measurements, while $y_{\text{ini}}$ represents the corresponding $T_{\text{ini}}$ output measurements. The vectors $u$ and $y$ represent future input and output trajectories over the prediction horizon $N$. The coefficient vector $g$ parameterizes these future trajectories as a linear combination of past input-output data stored in the Hankel matrices. This formulation constitutes the foundation of data-enabled predictive control, ensuring that feasible future trajectories are recoverable from historical measurements alone.

*2.4 Control Objective*

The objective of this research is to develop a control strategy that computes a sequence of control inputs $u = [u_k^\top \ u_{k+1}^\top \ \cdots \ u_{k+N-1}^\top]^\top$ at each time step $k$ so that the predicted output closely tracks a desired reference trajectory. In our formulation, the explicit, model-based description of the known subsystem (where the dynamics depend on the uncertain parameter $\theta$) is integrated with a data-driven constraint that captures the behavior of the unknown subsystem through Hankel matrix representations. The controller is designed by solving a robust optimization problem that minimizes a quadratic cost function:

$$J(u,y) = \sum_{i=0}^{N-1} \|y_{k+i} - r_{k+i}\|_Q^2 + \|u_{k+i}\|_R^2 \quad (5)$$

where $r_{k+i}$ is the reference trajectory and $Q$ and $R$ are positive definite weighting matrices, over the prediction horizon $N$, subject to both the model-based and data-driven system dynamics and the uncertainty constraints. This formulation is crafted to achieve robust performance by explicitly accounting for both parametric uncertainties in the known portion and stochastic disturbances, thereby ensuring reliable operation even under uncertain conditions.

In summary, this section formulates the Model-Inspired Distributionally Robust Data-enabled Predictive Control approach by partitioning the system into a known (model-based) subsystem and an unknown (data-driven) subsystem. It incorporates both parametric uncertainties via explicit models and stochastic disturbances through ambiguity sets and Hankel matrix representations, thereby establishing a robust foundation for predicting future system trajectories under uncertainty.

## 3. MDR-DeePC Framework

In this section, the framework of the Model-Inspired Distributionally Robust Data-enabled Predictive Control (MDR-DeePC) is presented, which integrates model-based and data-driven control strategies to manage systems with both known and unknown dynamics under parametric and stochastic uncertainties. This framework ensures robust stability and performance by leveraging the strengths of both approaches: the precision of model-based control for known dynamics and the flexibility of data-driven methods for unknown dynamics.

*3.1 Model-Inspired Control Strategy*

The Robust DeePC framework partitions the system into known and unknown subsystems, explicitly defining their dynamics and interactions. Consider a discrete-time system with states $x_k = [x_{k,k}^\top, x_{u,k}^\top]^\top$, where $x_{k,k} \in \mathbb{R}^{n_k}$ and $x_{u,k} \in \mathbb{R}^{n_u}$ represent the states of the known and unknown subsystems, respectively, $u_k \in \mathbb{R}^m$ is the control input, and $y_k \in \mathbb{R}^p$ is the measured output.

**Known Subsystem**: The known dynamics are given by the state-space model:

$$x_{k,k+1} = A_k(\theta)x_{k,k} + A_{ku}(\theta)x_{u,k} + B_k(\theta)u_k + w_{k,k} \quad (6)$$

where $A_k(\theta) \in \mathbb{R}^{n_k \times n_k}$, $A_{ku}(\theta) \in \mathbb{R}^{n_k \times n_u}$, and $B_k(\theta) \in \mathbb{R}^{n_k \times m}$ are parameter-dependent matrices, $\theta \in \Theta$ is an uncertain parameter, and $w_{k,k} \in \mathbb{R}^{n_k}$ is the process disturbance.

**Unknown Subsystem**: The unknown dynamics are captured implicitly through historical input-output data using Hankel matrices:

$$\begin{bmatrix} \mathcal{H}_L(u^d) \\ \mathcal{H}_L(y_u^d) \end{bmatrix} g = \begin{bmatrix} u_{\text{ini}} \\ y_{u,\text{ini}} \\ u \\ y_u \end{bmatrix} \quad (7)$$

where $u^d \in \mathbb{R}^{mT}$ and $y_u^d \in \mathbb{R}^{pT}$ are collected input-output data over $T$ time steps, $\mathcal{H}_L(\cdot)$ denotes the Hankel matrix of depth $L = T_{\text{ini}} + N$, $u_{\text{ini}}, y_{u,\text{ini}}$ are past trajectories, $u, y_u$ are future trajectories, and $g \in \mathbb{R}^{T-L+1}$ is a decision variable parameterizing the subsystem's behavior.

**Coupling**: The system output integrates both subsystems:

$$y_k = \begin{bmatrix} y_{k,k} \\ y_{u,k} \end{bmatrix} = \begin{bmatrix} C_k(\theta)x_{k,k} + D_k u_k \\ C_{uk}x_{k,k} + C_u x_{u,k} + D_u u_k \end{bmatrix} + v_k \quad (8)$$

where $C_k(\theta) \in \mathbb{R}^{p \times n_k}$ and $C_u \in \mathbb{R}^{p \times n_u}$ are output matrices, and $v_k \in \mathbb{R}^p$ is the measurement noise. The matrix $C_{uk} \in \mathbb{R}^{p \times n_k}$ captures the dependence of the data-driven outputs on the known state variables, allowing the structured coupling between modeled and unmodeled dynamics to be incorporated in the output equation.

Note that output measurements from the unknown subsystem are used to predict future outputs, and to approximate the unmeasured states through their coupling with the known subsystem dynamics. This interaction enables MDR-DeePC to account for the influence of the unknown state without requiring direct estimation. In particular, the effect of the unknown state $x_{u,k}$ on the evolution of the known state $x_{k,k+1}$ is approximated using the measured data-driven output $y_{u,k}$. The coupling term $A_{ku}x_{u,k}$ in the known dynamics is replaced with an expression $A_y y_{u,k}$, where the matrix $A_y$ is designed to capture the contribution of the unknown state encoded in $y_{u,k}$. This approximates the dynamics:

$$x_{k,k+1} = A_k(\theta)x_{k,k} + A_y(\theta)y_{u,k} + B_k u_k + w_{k,k} \quad (9)$$

This approximation maintains the structure of the hybrid model and ensures that all terms used in prediction and optimization are functions of measurable quantities.

## 3.2 Uncertainty Propagation

The MDR-DeePC framework addresses two distinct types of uncertainties:

**Parametric Uncertainties**: The parameter $\theta \in \Theta$ is treated as a deterministic uncertainty, where $\Theta$ is a compact set (e.g., a polytope). This affects the known subsystem matrices $A_k(\theta), A_y(\theta), B_k(\theta)$, and $C_k(\theta)$.

**Stochastic Uncertainties**: The disturbances $w_{k,k}$ and $v_k$ are stochastic with unknown distributions. To handle this uncertainty, data-driven ambiguity sets $\mathcal{P}_w$ and $\mathcal{P}_v$ are constructed using Wasserstein balls. These sets are built as follows: First, samples of $w_{k,k}$ and $v_k$ are estimated from historical input-output data, typically via residual analysis since direct measurements are unavailable. From these samples, empirical distributions $\mathbb{P}_w$ and $\mathbb{P}_v$ are formed, e.g., as discrete uniform distributions over the samples. The ambiguity sets are then defined as

$$\begin{aligned}\mathcal{P}_w &= \{\mathbb{P}: W_p(\mathbb{P}, \hat{\mathbb{P}}_w) \leq \epsilon_w\} \\ \mathcal{P}_v &= \{\mathbb{P}: W_p(\mathbb{P}, \hat{\mathbb{P}}_v) \leq \epsilon_v\}\end{aligned} \quad (10)$$

where $W_p$ is the $p$-Wasserstein distance, and $\epsilon_w, \epsilon_v > 0$ are radii tuned to balance robustness and performance (e.g., via cross-validation). These sets, constructed independently of $\theta$, ensure that the controller remains robust to any disturbance distribution within the specified proximity to the empirical distributions, as measured by the Wasserstein distance.

## 3.3 Distributionally Robust Optimization

Building on the control objective outlined in Section 2, this section presents the detailed formulation of the distributionally robust optimization problem. The goal is to minimize the cost function over a prediction horizon $N$, while ensuring robustness to both parametric uncertainties $\theta \in \Theta$ and stochastic disturbances $\xi = (w_k, v_k)$, where $w_k$ and $v_k$ represent process and measurement noise, respectively. Since $y$ depends on the control input $u$ and the disturbances $\xi$, the optimization problem is formulated with cost function defined in Eq. (5) as:

$$\min_u \sup_{\theta \in \Theta}\left(\sup_{\mathbb{P} \in \mathcal{P}} \mathbb{E}_{\mathbb{P}}[J(u, y(u, \xi))]\right) \quad (11)$$

subject to the following constraints:

(1) **Model-Based Constraints (Known Dynamics)**: As defined in Eqs. (6) and (8), which capture the evolution of the known states and their contributions to the measured output.

(2) **Data-Driven Constraints (Unknown Dynamics)**: As given in Eq. (7), based on Willems' fundamental lemma and historical input-output data.

(3) **Coupling and Feasibility Constraints**: The full system output and constraint sets are described by:

$$\begin{aligned}y_{k+i} &= \begin{bmatrix}y_{k,k+i}\\y_{u,k+i}\end{bmatrix},\\ u_{k+i} &\in \mathcal{U}, y_{k+i} \in \mathcal{Y}, \forall i = 0, \ldots, N-1\end{aligned} \quad (12)$$

with robustness enforced for all $\theta \in \Theta$ and all $\mathbb{P} \in \mathcal{P} = \mathcal{P}_w \times \mathcal{P}_v$.

The nested supremum in Eq. (10) poses computational challenges. To enhance tractability, a scenario-based approximation is adopted as follows:

$$\min_u \max_{i \in \{1,\ldots,M\}} \mathbb{E}_{\mathbb{P}_i}[J(u, y(u, \xi))] \quad (13)$$

where $M$ scenarios are generated by sampling $\theta_i \in \Theta$ and $\mathbb{P}_i \in \mathcal{P}$. This reformulation transforms the problem into a finite-dimensional robust optimization, which can be solved using standard convex programming tools such as CVX (Grant and Boyd, 2008) or YALMIP (Löfberg, 2004).

## 3.4 Data Validity Conditions

The data-driven representation of the unknown subsystem relies on the following assumptions.

**Assumption 1:** The input sequence $u^d$ is persistently exciting of order $L \geq T_{\text{ini}} + N + \mathbf{n}(\mathcal{B})$, where $\mathbf{n}(\mathcal{B})$ is the order of the unknown subsystem's minimal realization.

**Assumption 2:** The unknown subsystem is linear time invariant (LTI) during data collection.

These conditions ensure that the Hankel matrices $\mathcal{H}_L(u^d)$ and $\mathcal{H}_L(y_u^d)$ fully capture the subsystem's behavior, as per the Fundamental Lemma of Willems et al.

With the data validity conditions established, Algorithm 1 is presented to describe both the offline and online processes of the MDR-DeePC framework. This is accomplished by effectively combining model-based and data-driven constraints into a cohesive and robust optimization procedure.

---

**Algorithm 1: Compact MDR-DeePC Algorithm**

**Inputs**: Offline data $(u^d, y_u^d)$, initial horizon $T_{ini}$, prediction horizon $N$, uncertainty sets $\Theta, P_w, P_v$, terminal set $\chi_f$
**Outputs**: Control inputs $\{uk\}$ for each time step

**Offline (Once):**

1) Set $L = T_{ini} + N$; ensure $u^d$ is persistently exciting of order $L$
2) Construct Hankel matrices $\mathcal{H}_L(u^d)$ and $\mathcal{H}_L(y_u^d)$
3) Compute terminal set $\chi_f$

**for** each time step $k = 0,1,2,\ldots$ **do**

4) Measure (or estimate) current states $(x_{k,k}, x_{u,k})$
5) Formulate the **robust DeePC problem**:

   **Objective**: minimize cost $J(\cdot)$ from Eq. (5)
   subject to:
   -Model-based constraints: Eqs. (8) & (9)

   -Data-driven constraint: Eq. (7)

   -Robust feasibility: $u_{k+i} \in \mathcal{U}, y_{k+i} \in \mathcal{Y}, x_{k+N} \in \chi_f$
   Enforced for all $\theta \in \Theta, P \in P_w \times P_v$

6) Solve for $\{u_{k+i}\}_{i=0}^{N-1}$ and apply the first control $u_k$
7) **If adaptive**: update $P_w$ or $P_v$ with new data

**end for**

Note that step 7 ('if adaptive: update $P_w$ or $P_v$ with new data') specifies that if the controller is implemented adaptively, the ambiguity sets for the disturbances are updated online using newly acquired measurements, thereby keeping the controller's disturbance model up-to-date and ensuring robust performance under changing operating conditions. This algorithm bridges theory and practice, enabling real time implementation.

In general, the MDR-DeePC framework integrates model-based and data-driven control, addressing parametric uncertainties in $\theta$ and stochastic uncertainties in $w_{k,k}$ and $v_k$. Through explicit system formulation, separated uncertainty handling, rigorous data assumptions, and tractable optimization, it advances hybrid predictive control. Validation via a case study follows in Section 4.

## 4. Case Study

This section presents a case study demonstrating the practical effectiveness of the Model-Inspired Distributionally Robust Data-enabled Predictive Control (MDR-DeePC) framework. The objectives of this case study are:

**Hybrid Integration**: Showcasing the seamless integration of model-based control for known dynamics and data-driven predictive control for unknown dynamics.

**Robust Validation**: Illustrating the robustness of MDR-DeePC under significant parametric uncertainty and stochastic disturbances.

**Performance Demonstration**: Providing clear evidence of MDR-DeePC's tracking precision and vibration suppression capabilities.

### 4.1 Simulation Setup

A triple-mass-spring-damper system is considered as the simulation platform, where the first two masses are modeled with known dynamics, and the third mass represents the unknown dynamics subject to parametric variations. The simulation is designed to test the robustness of MDR-DeePC under both nominal operation and under a defined disturbance event.

For parametric uncertainties, the spring stiffness $k_3$ is set within $[80,120]$ N/m and the damping coefficient $c_3$ within $[3,7]$ Ns/m. For stochastic disturbances, process noise is given by $w(t) \sim \mathcal{N}(0, 0.1^2)$ and measurement noise by $v(t) \sim \mathcal{N}(0, 0.01^2)$.

The simulation runs for a total of 150 time steps, during which input–output data are collected under persistently exciting conditions. From these data, Hankel matrices are constructed with a depth $L = 24$ (with an initial horizon $T_{ini} = 4$ and a prediction horizon $N = 20$).

A key feature introduced in the simulation is the disturbance interval. This is a pre-defined segment of the simulation (specifically, from time steps 5 to 15) during which additional disturbances are injected into the system. The disturbance interval is used to evaluate the controller's ability to rapidly stabilize the system and maintain tracking performance under sudden changes.

The MDR-DeePC framework is executed in a receding horizon manner. At each time step, a robust optimization problem is solved—using CVX and the SDPT3 solver in MATLAB—with the most recent measurements and updated constraints. This online optimization incorporates the defined cost function (Eq. (5)) and ensures that the controller responds in real time, including applying integral action during the disturbance interval to reject the imposed disturbances.

### 4.2 Results and Discussion

The Model-Inspired Distributionally Robust Data-enabled Predictive Control (MDR-DeePC) demonstrated effective trajectory tracking and reliable stability in the triple-mass-spring-damper system. Figure 1 compares the control inputs, system outputs, and associated cost values achieved by MDR-DeePC (solid lines) and standard DeePC (dashed lines). This comparison illustrates the benefits of incorporating model knowledge and distributionally robust optimization, as MDR-DeePC exhibits improved robustness in handling parametric uncertainties and stochastic disturbances compared to standard DeePC, which relies solely on data-driven methods without explicit uncertainty handling. The shaded region (time steps 5–15) represents the disturbance interval with significant parametric uncertainties and stochastic disturbances.

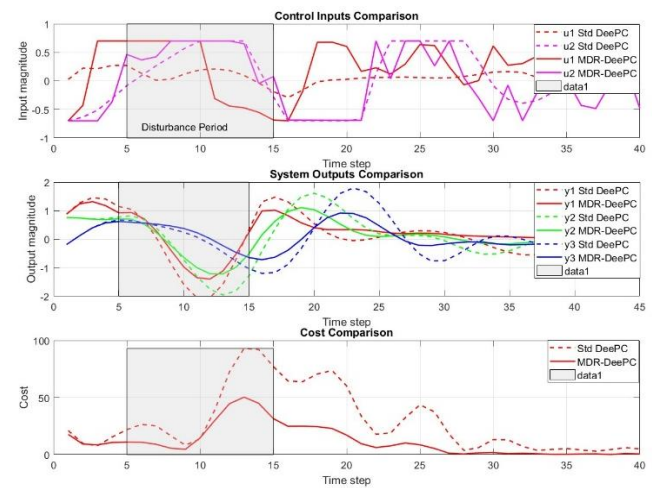

Fig. 1. Comparison of MDR-DeePC and Standard DeePC on Control Inputs, System Outputs.

System outputs ($y_1$, $y_2$, $y_3$) under MDR-DeePC exhibited substantially reduced deviations and oscillations relative to standard DeePC. During disturbances, MDR-DeePC maintained tracking errors consistently below 5% of the desired reference trajectories and stabilized outputs more rapidly. In contrast, standard DeePC showed larger oscillations, higher deviations, and slower convergence to reference values, indicating reduced robustness under similar conditions.

The control input trajectories illustrate that MDR-DeePC actively adjusted the inputs with pronounced variations to quickly counteract disturbances, leading to less smooth control signals compared to standard DeePC. Standard DeePC generated smoother input profiles but failed to achieve effective stabilization, resulting in sustained oscillations and higher output deviations. The corresponding cost plot quantitatively supports this observation: MDR-DeePC achieved lower peak costs during disturbances and more rapid reduction afterward, highlighting improved performance in managing disturbances and uncertainties.

MDR-DeePC's robustness advantage stems from the explicit use of distributionally robust optimization coupled with model-based constraints, enabling dynamic adaptation to persistent disturbances. The adaptive weighting mechanism further optimized the trade-off between input variations (control effort) and precise output tracking.

Table 1. Quantitative Performance Metrics Comparison between Standard DeePC and MDR-DeePC

|  | Std DeePC | MDR-DeePC | Improvement |
|---|---|---|---|
| Total cost | 1152.6538 | 475.5094 | 58.75% |
| Maximum output deviation | 7.7705 | 5.1638 | 33.55% |
| Settling time (steps) | 25 | 25 | 0.00% |
| Peak-to-Peak amplitude | 3.2005 | 2.3289 | 27.23% |

A quantitative comparison of performance metrics between standard DeePC and MDR-DeePC is summarized in Table 1. MDR-DeePC achieves a 58.75% reduction in total accumulated cost and a 33.55% reduction in maximum output deviation compared to standard DeePC. Although both controllers exhibit identical settling times (25 steps), MDR-DeePC achieves a 27.23% reduction in output peak-to-peak amplitude, indicating more effective suppression of oscillations. These results confirm that MDR-DeePC improves overall tracking precision and control efficiency under uncertainty.

## 5. CONCLUSIONS

This paper introduced the Model-Inspired Distributionally Robust Data-enabled Predictive Control (MDR-DeePC) framework, effectively integrating model-based and data-driven methodologies within a unified distributionally robust optimization structure. MDR-DeePC was specifically designed to address both parametric uncertainties and stochastic disturbances in hybrid dynamical systems, aiming to provide robustness while maintaining high tracking performance and vibration suppression capabilities.

Through a detailed case study on a triple-mass-spring-damper system, MDR-DeePC demonstrated notable advantages over traditional predictive control methods, including precise trajectory tracking, effective disturbance rejection, and excellent operational robustness. The adaptive weighting scheme within the optimization strategy ensured balanced performance under varying uncertainty conditions, and the computational complexity remained manageable for real-time implementations.

Future work will explore adaptive or automated tuning strategies for robust optimization parameters to reduce the offline calibration effort. Extending MDR-DeePC to accommodate nonlinear and time-varying systems represents another promising direction, potentially broadening its applicability across diverse, complex control scenarios.